\documentstyle[sprocl,psfig,epsf]{article} 

\def\Journal#1#2#3#4{{#1} {\bf #2}, #3 (#4)}

\def\NPA{{\em Nucl. Phys.} A}
\def\NPB{{\em Nucl. Phys.} B}

\def\PLB{{\em Phys. Lett.} B}

\def\PRL{\em Phys. Rev. Lett.}

\def\PRD{{\em Phys. Rev.} D}
\def\PRC{{\em Phys. Rev.} C}

\def\be{\begin{equation}}
\def\ee{\end{equation}}
\def\bea{\begin{eqnarray}}
\def\eea{\end{eqnarray}}
\newcommand{\beq}{\begin{equation}}
\newcommand{\eeq}{\end{equation}}
\newcommand{\beqa}{\begin{eqnarray}}
\newcommand{\eeqa}{\end{eqnarray}}

\begin{document}

\hfill {{\small FZJ-IKP(TH)-1998-26}} 

\smallskip   

\title{CHIRAL DYNAMICS - STATUS AND PERSPECTIVES 
\footnote{Plenary talk at BARYONS~98,
Bonn, Germany, September 1998.}} 

\author{Ulf-G. Mei{\ss}ner}

\address{FZ J\"ulich, IKP (Th), 
D-52425 J\"ulich, Germany\\E-mail: Ulf-G.Meissner@fz-juelich.de}

\maketitle\abstracts{I review the status of chiral perturbation
theory  in the one--nucleon sector and give some predictions
to be tested. I then discuss various methods to go to higher
energies (inclusion of the $\Delta (1232)$, dispersion relations)
and also to higher precision (virtual photons and isospin violation).
}


\section{Effective field theory of QCD}

\noindent In the sector of the three light quarks $u$, $d$ and $s$, 
QCD admits a global chiral symmetry softly broken by the quark mass term,
\begin{equation}
{\cal H}_{\rm QCD} = {\cal H}_{\rm QCD}^0  +  {\cal H}_{\rm QCD}^{\rm SB}
=  {\cal H}_{\rm QCD}^0  + 
 m_u \, \bar u u + m_d \, \bar d d + m_s \, \bar s s \,\, ,
\end{equation}
where "light" means that the current quark mass at a renormalization
scale of $\mu = 1\,$GeV can be treated as small compared to the
typical scale of chiral symmetry breaking, $\Lambda_\chi \simeq 4 \pi
F_\pi \simeq 1.2\,$GeV, with $F_\pi \simeq 93\,$MeV the pion decay
constant. The tool to investigate these issues is chiral perturbation
theory (CHPT). In CHPT, the basic degrees of freedom are
the Goldstone boson fields coupled to external sources and matter
fields, like e.g. the nucleons. QCD is mapped onto an effective
hadronic Lagrangian formulated in terms of these asymptotically observed 
fields. Any matrix element involving nucleons, pions, photons and so on can
be classified according to its {\it chiral dimension}, which counts the
number of external momenta, quark mass insertions and inverse powers
of heavy mass fields. Denoting these small parameters collectively as
$p$, CHPT allows for a systematic perturbative expansion in powers of
$p$, with the inclusion of loop graphs and local terms of higher
dimension. The latter are accompanied by a priori unknown coupling
constants, the so--called low--energy constants (LECs). This is the
so--called {\it chiral} expansion, which is nothing but an energy
expansion reminiscent of the ancient Euler--Heisenberg treatment of
light--by--light scattering in QED at photon energies much smaller
than the electron mass. In QCD, the equivalent heavy mass scale is essentially
set by the first non--Goldstone resonances, i.e. the $\rho , \omega$ mesons.
This dual
expansion in small momenta and quark masses can be mapped one--to--one onto
an expansion in powers of Goldstone boson loops, where an $N$--loop graph
is suppressed by powers of $p^{2N}$.\cite{wein79} The leading terms are
in general tree graphs with lowest order insertions leading to the
celebrated current algebra (CA) results. While in the meson sector one only
has terms with even powers of $p$, in the baryon sector terms with odd
and even powers in $p$  are allowed, so that for example a complete
one--loop calculation includes terms of order $p^3$  {\it and}  $p^4$.
In  what  follows, I will mostly be concerned with  the two--flavor
sector, i.e. the pion--nucleon system.
For a detailed review, I refer to ref.\cite{bkmrev}

\section{A short status report}
I first  consider a variety of predictions in comparison to the existing
data and then discuss some predictions to be tested. Pion--nucleon scattering
and pion induced pion production are the classic testing grounds for baryon
CHPT  and precise data have become available over the last years. In fig.1
(left panel), the CHPT  predictions for the S--wave  $\pi N$ scattering
lengths\cite{FMS} are shown in comparison to the beautiful measurements
of  the level shifts of pionic hydrogen and deuterium performed at PSI.
Most  important is the observation that the shift from the CA point
to the CHPT square is largely a pion loop effect, i.e. one sensitively
probes the chiral structure of QCD.  
\begin{figure}[h]
\vspace{6.5cm}
\includegraphics{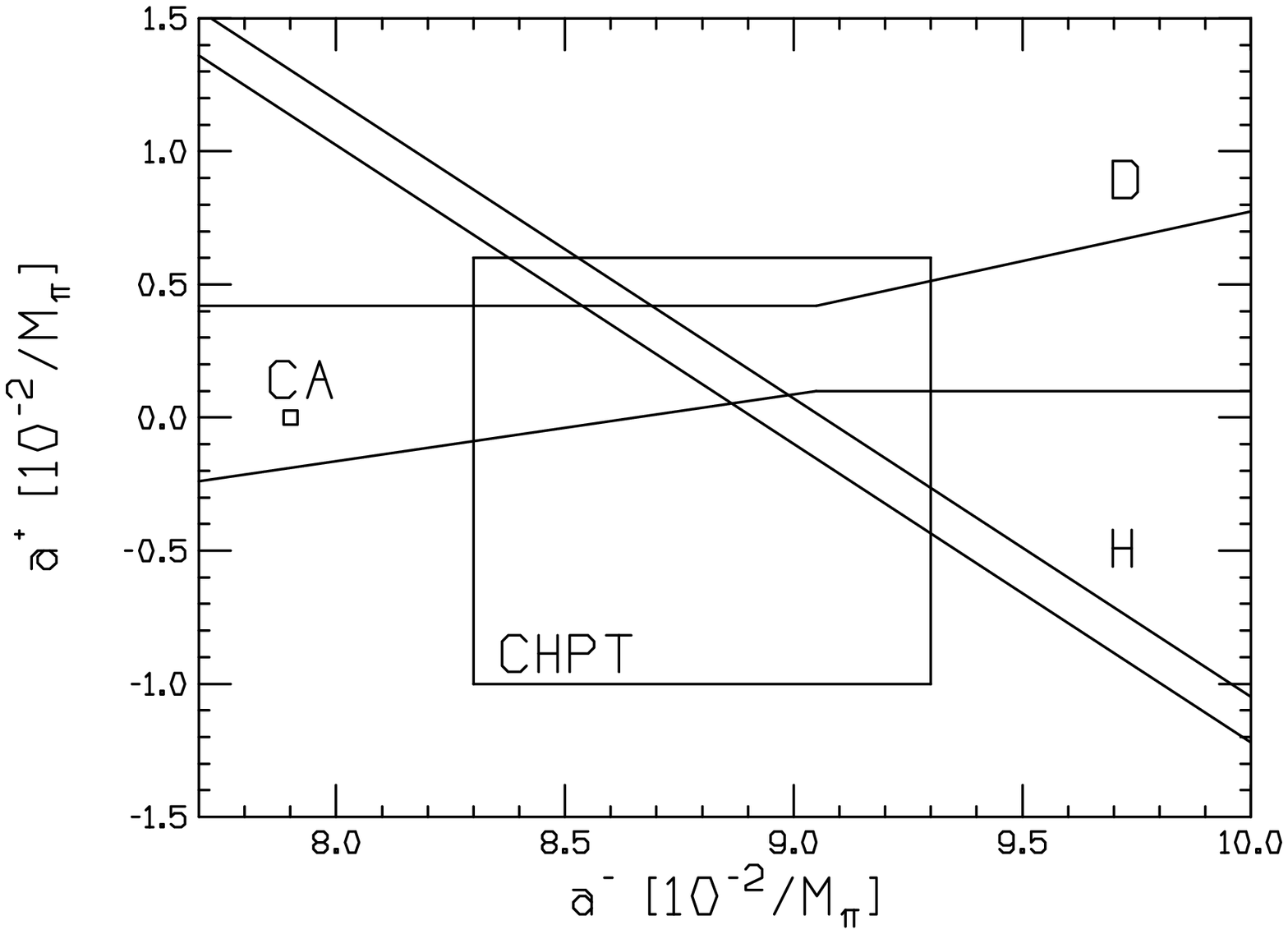}
\hspace{5.cm}
\includegraphics{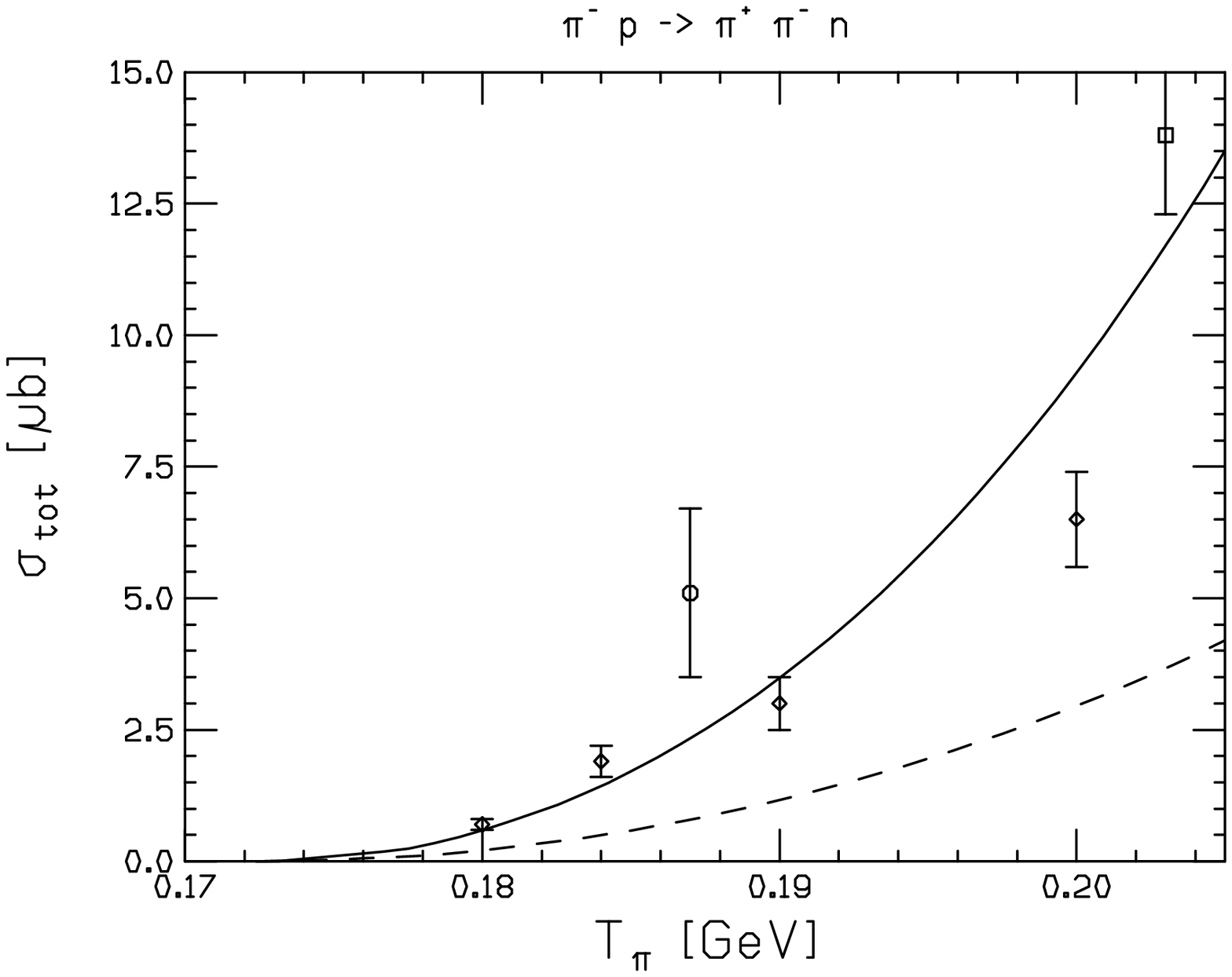}
\vspace{-2.2cm}
\caption{Left panel: CHPT prediction for the S--wave scattering 
lengths in comparison
to the pionic hydrogen (H--band) and deuterium (D--band) data  from PSI.
Right panel: Prediction for $\pi^- p \to \pi^+ \pi^- n$ at leading (dashed)
and next--to-leading (solid line) order. The new TRIUMF data (at $T_\pi =
0.18,0.184,0.19,0.20\,$GeV) were obtained after the CHPT prediction.}
\end{figure}
In the right panel of fig.1, a prediction  based on a next--to--leading
(NLO) calculation for $\pi^- p \to \pi^+ \pi^- n$ is shown in comparison
to the two older data points (large error  bars) and  the new ones from
TRIUMF  (small error bars) obtained after the prediction. Note that
the agreement shows the expected pattern: The closer one is to threshold,
the better the agreement. More details on these topics  are given by
Martin Sevior.\cite{MS}
Much experimental and theoretical activity has been 
focused on photo--nucleon reactions. In particular, neutral pion production off
the proton and off deuterium has been measured very precisely. A typical
result is shown in the left panel of fig.2, based on the calculation of
ref.\cite{bkmpi0l}. Clearly, one observes the unitary cusp at the $\pi^+ n$
threshold and the small value of the electric dipole amplitude at the
$\pi^0 p$ threshold is mostly due to the so--called triangle diagram.
Furthermore, these data clearly rule out  the ``NOLET'' by many  standard
deviations.\cite{gerulf} Note, however, that the convergence of the
chiral expansion in this case is not very  rapid and better tests are
indeed given in the P--waves, as discussed by Thomas Walcher.\cite{TW}
\begin{figure}[htb]
\vspace{6.cm}
\includegraphics{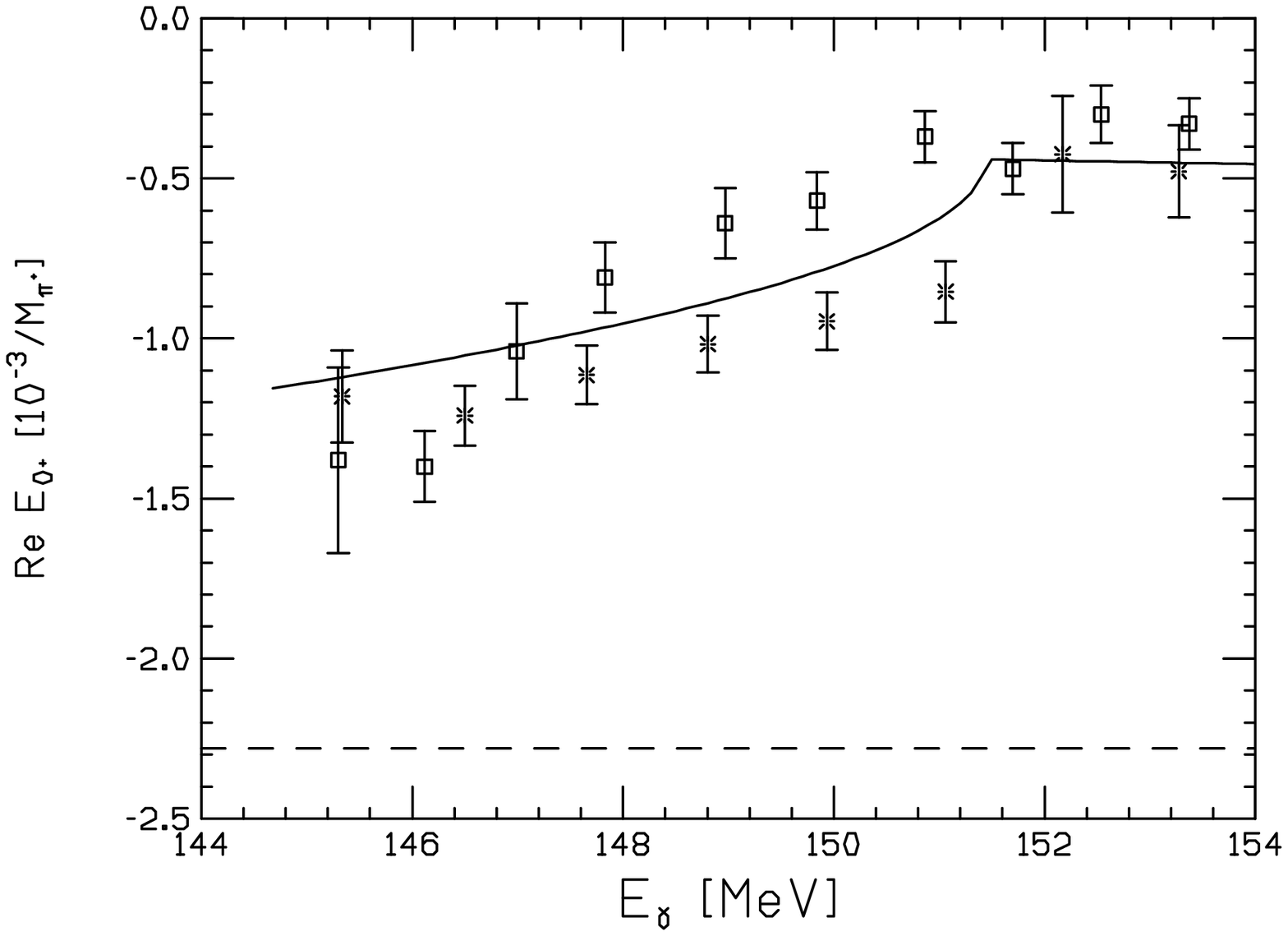}
\hspace{4.8cm}
\includegraphics{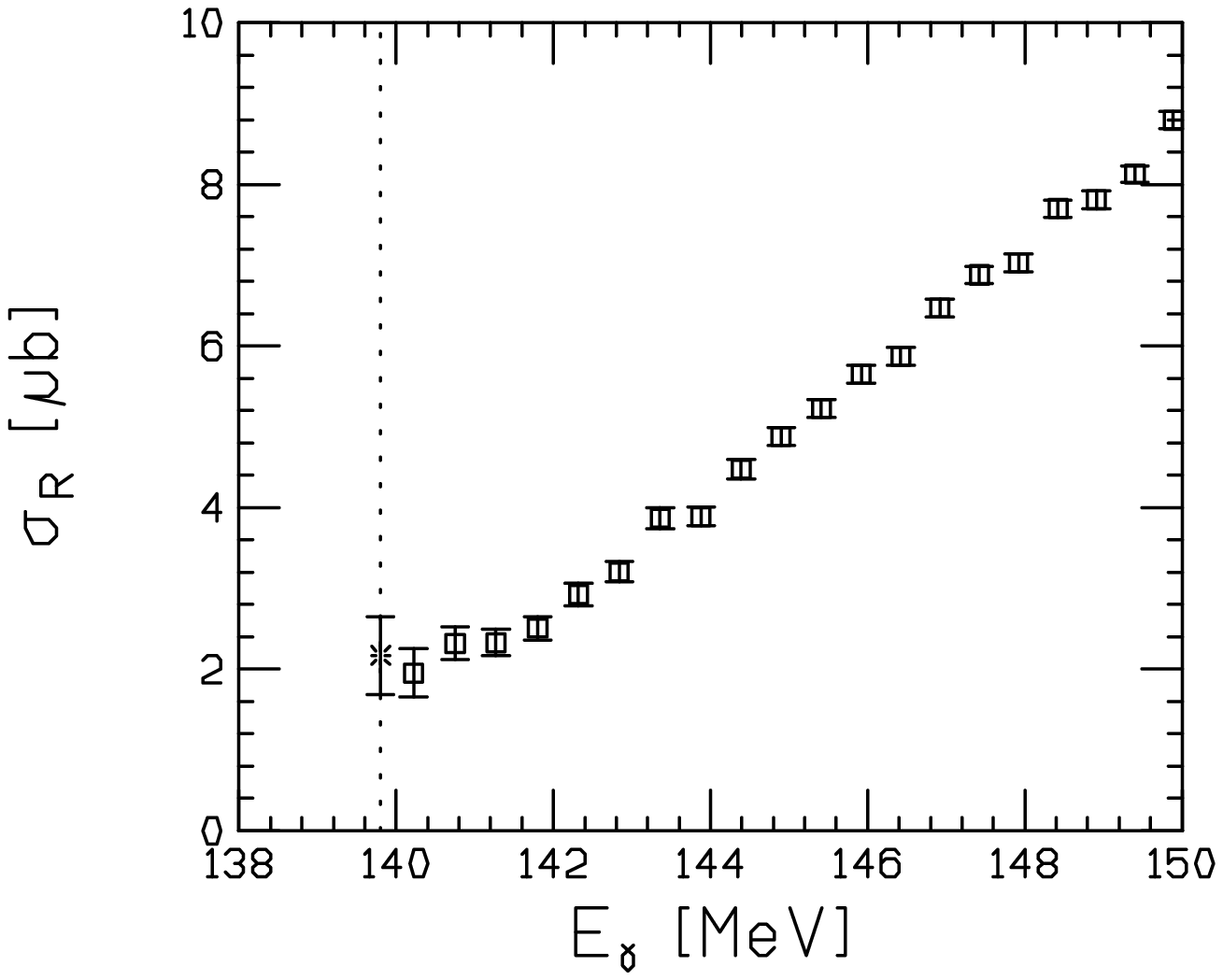}
\vspace{-2.2cm}
\caption{Left panel: CHPT prediction (solid line) 
for the electric dipole amplitude
in   $\gamma p \to \pi^0 p$ in comparison to the  data  from MAMI and SAL.
The dashed line  gives the prediction of the ancient ``NOLET''.
Right panel: Reduced total cross section for $\gamma d \to \pi^0 d$. The
data from SAL are depicted by the boxes, the CHPT threshold prediction is the
star on the the dotted line (indicating the threshold photon energy).} 
\end{figure}
To my  opinion, a particularly nice example of chiral dynamics at
work is the prediction of the electric dipole amplitude for neutral
pion production off deuterium,\cite{silas} which  is  very sensitive
to the elementary neutron amplitude. The chiral prediction for the
deuteron is $E_d = (-1.8\pm 0.2)\cdot 10^{-3}/M_{\pi}$, only
20\% above the empirical value from SAL,\cite{berg} $E_d^{\rm emp}
 = (-1.45\pm 0.09)\cdot 10^{-3}/M_{\pi}$ and overlaps within 1.5~$\sigma$.
The corresponding reduced total cross section (after subtraction of the
breakup channel) measured at SAL is shown in fig.2 (right panel)
 together with the CHPT prediction.
This shows that the elementary neutron amplitude is in fact as {\it large}
as predicted by CHPT and at total variance with the ``NOLET'' prediction,
$E_{0+}^{\pi^0 n} \simeq 0$.
One can also extend these calculations to SU(3). While  there are many
problems due to the large kaon mass and (subthreshold) resonances, in
some cases like the magnetic moments or the recoil polarization in
$\gamma p \to K \Lambda$ the conventional framework is useful, see
e.g. ref.\cite{SU} I will come back to this later on. 
Of course, there are also problems. For example, the recent TRIUMF
measurement of radiative  muon capture
\cite{TRImu} can be used to
infer the induced pseudoscalar coupling  constant. The extracted value
is about 50\% larger than  the very precise CHPT prediction.\cite{bkmgp}
At present, this discrepancy has not been resolved.\cite{THp} A more
detailed status report of chiral dynamics (including also the meson sector)
can be found  in refs.\cite{Ecker,ulfhugs}
To end this section, I turn to some predictions which  have not been
tested. The importance of looking at spin--dependent Compton scattering
was first pointed out in ref.\cite{bkkm} and detailed predictions for
the so--called spin--polarizabilities can be found in ref.\cite{Th1}.
From these four structure constants, $\gamma_1$and $\gamma_3$ are
particularly sensitive to the chiral  pion loops. In virtual Compton
scattering, one can measure the so--called generalized polarizabilities,
as discussed by Nicole d'Hose.\cite{nicole} In fig.3, I show some
genuine predictions due to Hemmert et  al.\cite{Th2,Th3,Th4}
Of special interest is the generalized magnetic polarizability $\beta (q^2)$
since its positive slope at $q^2 =0$ is a {\it unique} chiral prediction.
\begin{figure}[h]
\vspace{7cm}
\includegraphics{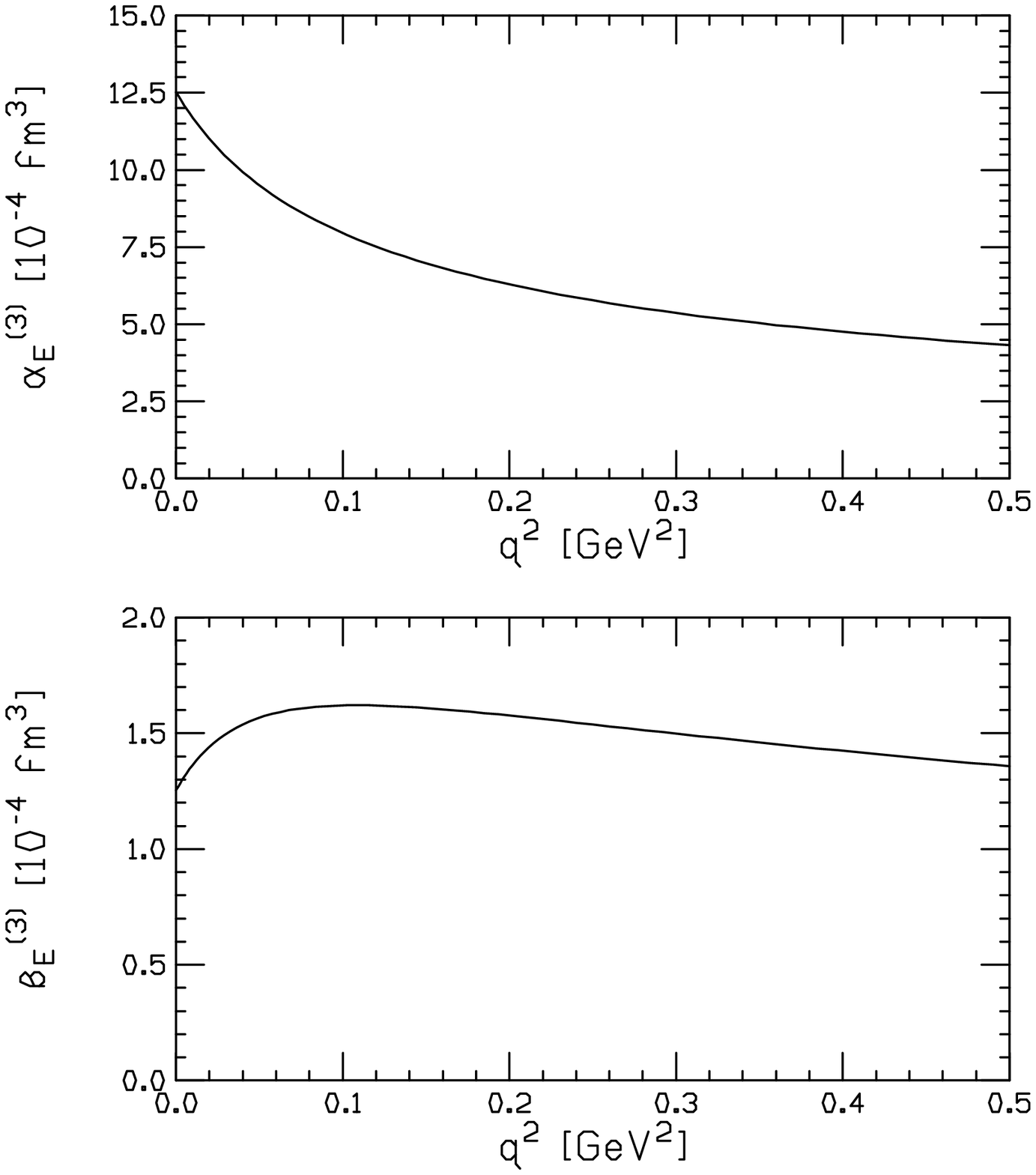}
\hspace{5.cm}
\includegraphics{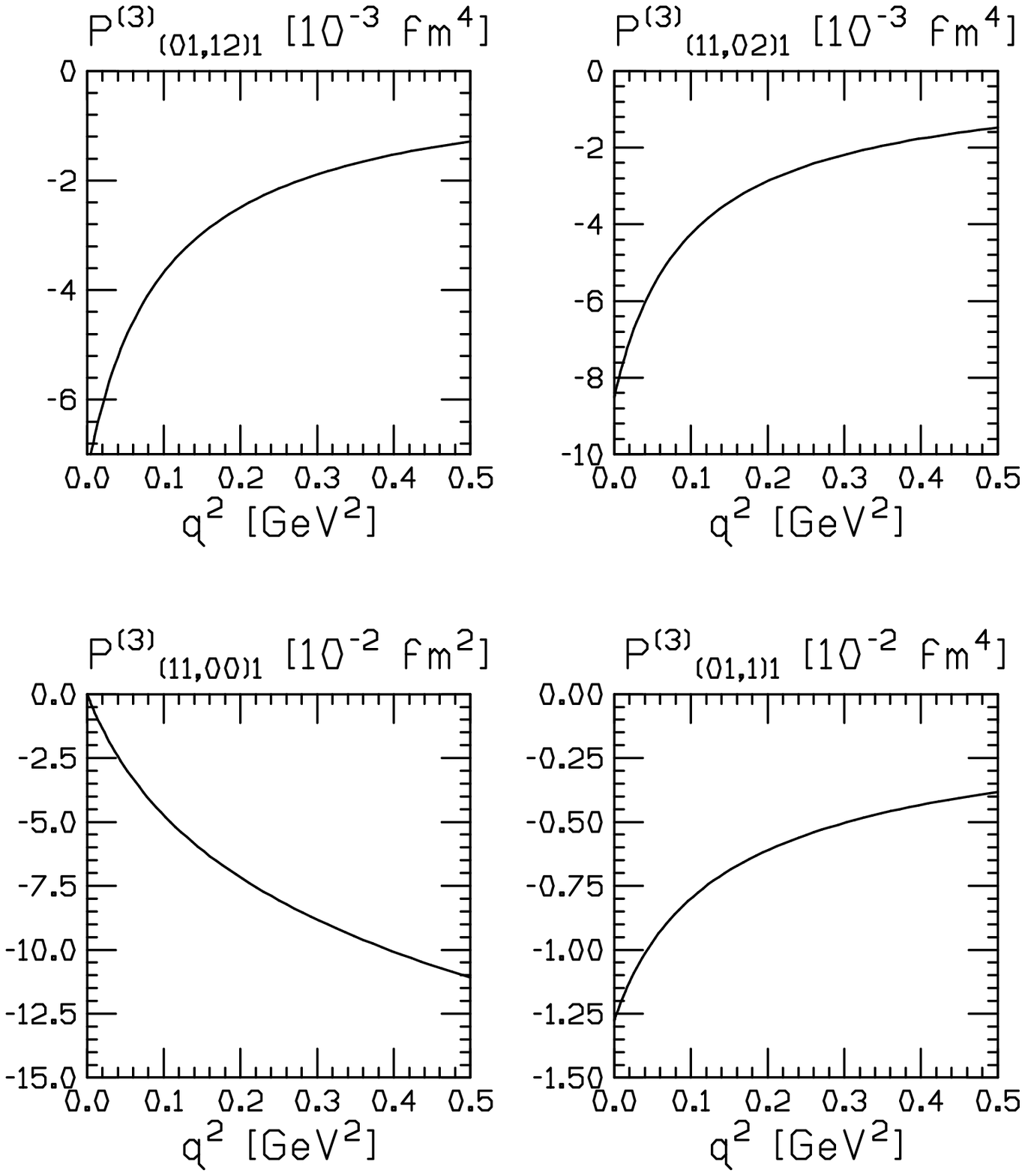}
\vspace{-1.5cm}
\caption{Left panel: CHPT prediction for the generalized spin--independent
polarizabilities as a function of the squared momentum transfer. 
Right panel: $p^3$ prediction for the four independent spin--dependent
generalized polarizabilities as a function of the squared momentum transfer.
} 
\end{figure}
Also, the P--wave low--energy theorems (LETs) for pion 
electroproduction\cite{bkmlet}
have so far only been used in the analysis of the data on $\gamma^\star p \to
\pi^0 p$ measured at NIKHEF and MAMI. More complete angular distributions
and polarization observables will have to be measured to uniquely extract
these P--waves and test the LETs. These are only  a few examples of the
richness of chiral predictions in the respective threshold regions which
pose stringent tests on our understanding of chiral symmetry in QCD.

\section{Perspectives}
So far, I have considered the effective pion--nucleon field theory.
It is most  predictive in threshold situations. Naturally, one would
like to extend the scheme to higher energies. I discuss here a few
possibilities to do that. Also, to address certain questions one still needs
a more refined and precise machinery  even in the $\pi N$ sector as I
will argue below. First, however, let me describe some attempts
to extend the range of applicability of the EFT.

\subsection{Inclusion of the $\Delta$} 
Although the inclusion of the decuplet was originally formulated for
SU(3),\cite{jmd} let us focus here on the simpler two--flavor
case. Among all the resonances, the $\Delta (1232)$ plays a particular
role for essentially {\it two} reasons. First, the $N\Delta$ mass
splitting is a small number on the chiral scale of 1 GeV,
$\Delta \equiv m_\Delta - m_N = 293 \, {\rm MeV} \simeq 3 F_\pi$,
and second, the couplings of the $N\Delta$ system to pions and photons
are very strong, much stronger than any other nucleon resonance.
So one could consider $\Delta$ as a small parameter. It is, however,
important to stress that in the chiral limit, $\Delta$ stays finite
(like $F_\pi$ and unlike $M_\pi$). Inclusion of the spin--3/2 fields
like the $\Delta (1232)$ is therefore based on phenomenological
grounds but also supported by large--$N_c$ arguments since in that
limit a mass degeneracy of the spin--1/2 and spin--3/2 ground state
particles appears. Recently, Hemmert, Holstein and Kambor\cite{HHK}
proposed a systematic way of including the $\Delta (1232)$ based on an 
effective Lagrangian of the type ${\cal L}_{\rm eff} [U, N , \Delta ]$
which has a systematic ``small scale expansion'' in terms of {\it
  three} small parameters (collectively denoted as $\epsilon$). These
are
$\frac{E_\pi}{ \Lambda}$,
$\frac{M_\pi}{\Lambda}$ and
$\frac{ \Delta}{\Lambda}$,
with $\Lambda \in [ M_\rho, m_N, 4\pi F_\pi ]$. Starting from the
relativistic pion--nucleon-$\Delta$ Lagrangian, one writes the
nucleon ($N$) and the Rarita--Schwinger  ($\Psi_\mu$)
fields  in terms of velocity eigenstates (the nucleon four--momentum
is $p_\mu = m v_\mu + l_\mu$, with $l_\mu$ a small off--shell 
momentum, $v \cdot l \ll m$ and similarly for the $\Delta (1232)$,
$N = {\rm e}^{-imv \cdot x} \, (H_v + h_v) \,$, 
$\Psi_\mu = {\rm e}^{-imv \cdot x} \, (T_{\mu \,v} + t_{\mu \,v})\,)$,
and integrates out the ``small'' components $h_v$ and $ t_{\mu \,v}$
by means of the path integral formalism developed in Ref.\cite{bkkm}
The corresponding heavy baryon effective field theory 
in this formalism does not only
have a consistent power counting but also $1/m$ suppressed vertices
with fixed coefficients that are generated correctly (which is much simpler
than starting directly with the ``large'' components and fixing these
coefficients via reparametrization invariance). Since the spin--3/2
field is heavier than the nucleon, the residual mass difference 
$\Delta$ remains in the spin--3/2 propagator and one therefore has to
expand in powers of it to achieve a consistent chiral power counting.
The technical details how to do that, in particular how to separate
the spin--1/2 components from the spin--3/2 field, are given in 
ref.\cite{HHK} Let me now consider two examples.
The first one is related to the
scalar nucleon form factor (for an early calculation
capturing the essence of  the small scale expansion, see ref.\cite{bkmzm}).
The scalar form factor is given  by
\begin{equation}
\sigma_{\pi N} (t) = \frac{1}{2} (m_u+m_d) \, \langle p' \, | \bar{u}u
+ \bar{d}d \,| p  \, \rangle \,\, , \quad t = (p' -p)^2 \,\, .
\end{equation}
Of particular interest in the analysis of $\sigma_{\pi N}$ is the
Cheng--Dashen point, $t = 2M_\pi^2 \, , \, \nu=0$ (at this unphysical
kinematics, higher order corrections in the pion mass  are the
smallest) and one evaluates 
$\Delta \sigma_{\pi N} \equiv \sigma_{\pi N} (2M_\pi^2) - \sigma_{\pi
 N} (0)\,$.
To one loop and order ${\cal O}(\epsilon^3)$, $\Delta \sigma_{\pi N}$
is free of counter terms and just given by two simple 
one loop diagrams (one with  an intermediate $N$ and the other with
an intermediate  $\Delta$)\cite{bkmzm,JK} 
\begin{eqnarray}
\label{CDres}  
\Delta \sigma_{\pi N} &=& \frac{3 g_A^2 M_\pi^3}{64 \pi F_\pi^2} +
\frac{g^2_{\pi N \Delta}}{6\pi^2 F_\pi^2} \frac{M_\pi^4}{\Delta}
\biggl\{\frac{5}{18}- \frac{\pi}{24} + \frac{5}{6} \ln
\frac{2\Delta}{M_\pi} + \dots \biggr\} \nonumber \\ 
&=& (7.4 + 4.1) \, {\rm MeV} = 11.5 \, {\rm MeV} \,\, ,
\end{eqnarray} 
where the first term comes from the intermediate nucleon and scales
as $M_\pi^3$ whereas the other terms come from the loop graph with
the intermediate $\Delta (1232)$ and scale as $M_\pi^4 / \Delta$,
which are therefore both ${\cal O}(\epsilon^3)$. In the chiral
expansion of QCD, however, the first term is ${\cal O}(p^3)$ while the
second is of order $p^4$. Therefore, the epsilon expansion can be considered as
a rearrangement of the chiral expansion.
The correction due to the delta goes in the right direction, 
but the resulting number is still about 30\% below the one of
the  dispersion--theoretical analysis (supplemented by chiral 
symmetry constraints) of Gasser, Leutwyler and Sainio, $\Delta
\sigma_{\pi N} = (15 \pm 1)\,$MeV.\cite{gls} 
This points towards the importance of higher order 
effects, as I will discuss in more detail below.
The interplay between the chiral and the small scale expansion has been studied
in ref.\cite{BFHM} There, the nucleons' electroweak form factors are considered.
These calculations also serve as a first exploratory study of renormalization
and decoupling within the small scale expansion. In particular, it is argued
that the pion--nucleon effective theory in the presence of the $\Delta$ differs
from the ``pure'' $\pi N$ EFT, in particular, the $\beta$--functions related to
various dimension three operators needed for the renormalization obtain additional
terms in the EFT with explicit deltas. As an  example, consider the 
electromagnetic nucleon structure as parametrized by the electroweak formfactors (ff).
To third order in the chiral expansion, the isovector Pauli ff can be predicted without
any LEC, whereas the expression for the Dirac ff contains one LEC. This can be adjusted
to give the empirical isovector charge radius. The numerical value of this LEC differs,
of course, in the chiral and in the small scale expansion. The predictions of the
chiral and the small scale expansion are consequently indistinguishable 
for $F_1^V (q^2)$, cf. fig.4. For the isovector Pauli ff,
\begin{figure}[htb]
\hspace{4.cm}
\epsfysize=1.6in
\epsffile{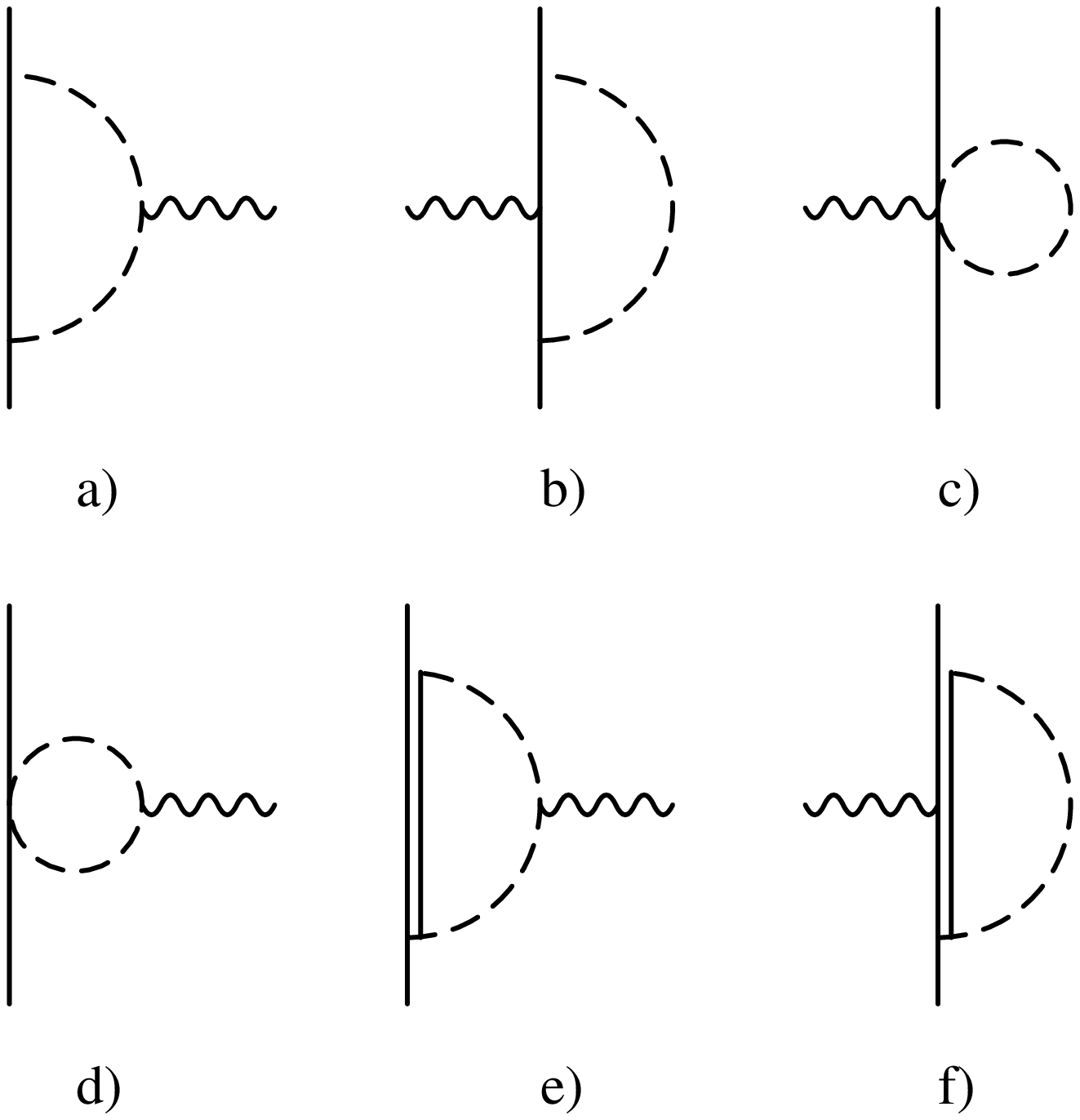}
\vspace{-1.7truein}
\epsfysize=1.7in
\epsffile{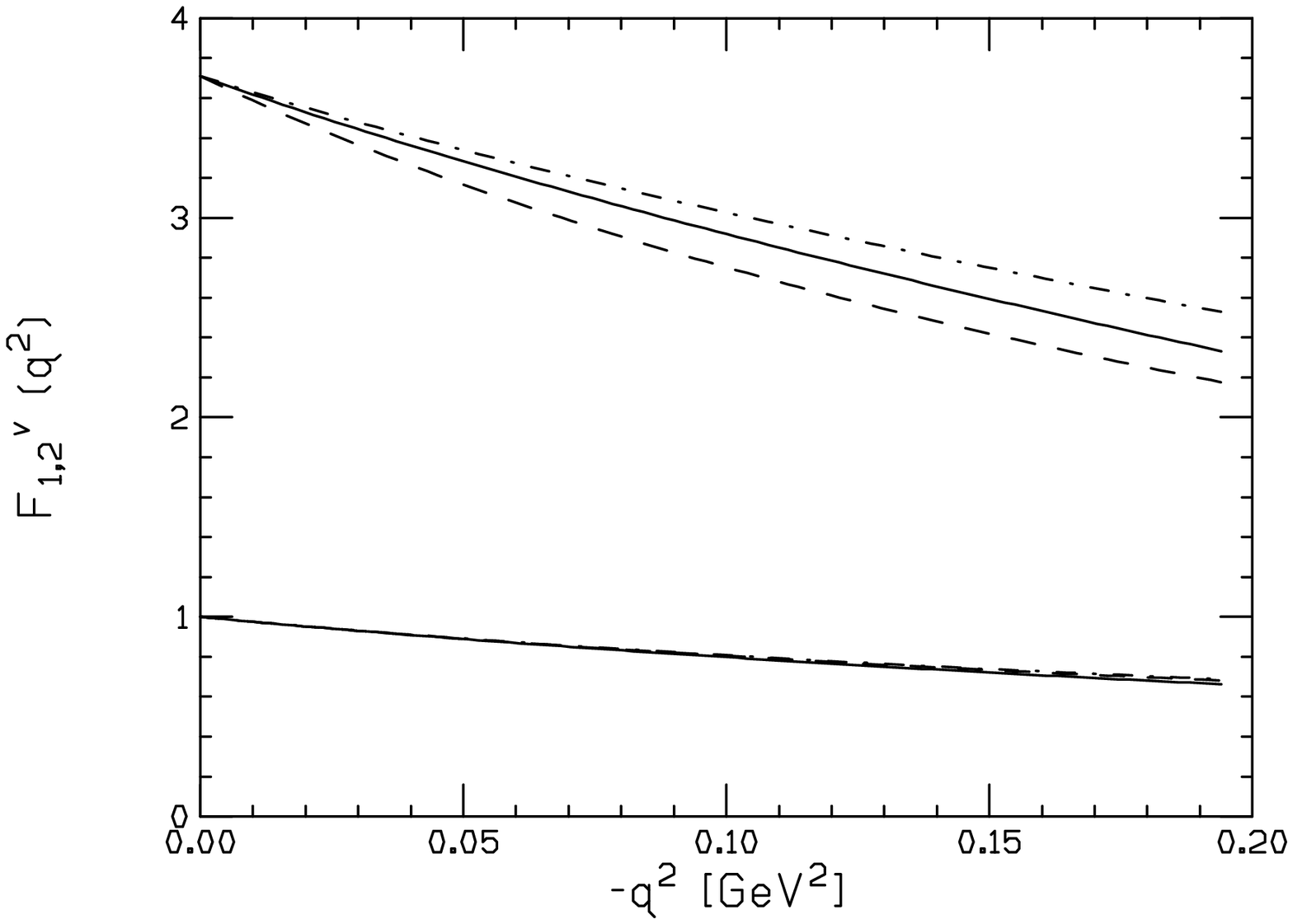}
\caption{
Left panel: The nucleon isovector Dirac (lower curves) and Pauli (upper curves)
form factors. The prediction of the small scale (chiral) expansion
to order $\epsilon^3$ ($p^3$) is shown by the solid (dot-dashed) lines.
The data are represented by the result of a 
dispersion--theoretical analysis (dashed lines). Right panel: One loop
graphs with intermediate nucleons (solid lines) and deltas (double lines).
}
\end{figure}
the novel graphs with intermediate deltas 
(see right panel in fig.4) improve the prediction of the radius and lead
to a better description of the ff at small and moderate momentum transfer, see. fig.4.
Of course, this EFT is supposed to work best in the $\Delta$ region, see e.g. the
talk by Gellas\cite{GC} on the $\Delta \to N\gamma$ transition form factors and the
related work on pion photoproduction in the $\Delta$ region.\cite{BHM} Furthermore,
there has been sizeable activity in computing real and virtual Compton scattering
in the framework of the epsilon expansion, see e.g.\cite{Th1,Th4} What is still missing
is a complete renormalization of the generating functional to third order and some
explicit calculations to ${\cal O}(\epsilon^4)$, to compare e.g. with the precise
chiral predictions for the nucleons' em polarizabilities.

\subsection{Marriage with dispersion relations}

Another possibility of extending the range of applicability of CHPT
(or to sharpen the chiral predictions at low energies) is the
use of dispersion relations (DR), 
for some early work in the meson sector
see e.g.\cite{tnt,gm} (for a more pedagogical introduction, see e.g. ref.\cite{JD}). 
Let me first make some general remarks.
A generic amplitude for some process can be written in terms of an n--times
subtracted dispersion relation (where  the number of subtractions
depends on the convergence properties and generally is small)
\beq
A(t) = \sum_{i=0}^{n-1} a_i \, t^i + \frac{t^n}{\pi} \int_{t_0}^\infty
\frac{dt'}{{t'}^n} \frac{{\rm Im\,} A(t)}{t-t' -i\epsilon}~.
\eeq
These subtraction constants $a_i$ play a role similar to the LECs in the
corresponding chiral calculation and thus chiral symmetry can be
used to constrain them. Similar remarks hold for the spectral function
Im~$A(t)$ at low $t$. Of course, there is no free lunch - precise cross
section data at intermediate and large momenta are needed to make DRs a 
viable and accurate tool (note that sometimes the large momentum
behaviour is modeled giving rise to some unwanted model--dependence).
Note also that the analytic structure in the
two approaches is the same (as demanded by analyticity). 
In the dispersive approach, one essentially
sums up classes of loop diagrams (the so--called unitarity corrections)
allowing one to obtain a precise representation at low and higher
energies. What the dispersive treatment can in general not provide is
linking Green functions of different processes (as it is the case
in CHPT) and only under special circumstances the enhancement of
certain LECs due to IR logarithms can be unraveled.\cite{gm}
Clearly, the combination of dispersion theory with CHPT  constraints
is a powerful tool which has not yet been explored in big detail in
the nucleon sector. One exception to this is the work on the
scalar form factor in ref.\cite{gls} Since the nucleon is a 
composite object, $\sigma_{\pi N} (t)$ must fulfill a once--subtracted
DR. The corresponding spectral function reads 
\beq
{\rm Im\,}\sigma_{\pi N} (t) = \frac{3}{2} \frac{\Gamma_\pi^* (t) \,
f_+^0 (t)}{4m^2 - t} \sqrt{1 - \frac{4M_\pi^2}{t}}~,
\eeq
with $\Gamma_\pi$ the scalar form factor of the pion,
$\langle \pi^a (p')|\hat{m}(\bar u u +\bar d d)|\pi^b (p)\rangle
\sim \Gamma_\pi (t)$ and $f_0^+ (t)$
the $I=J=0$ $\pi N$ partial wave (analytically continued from the
existing data). The evaluation of the dispersive integral leads
to the abovementioned result for $\Delta \sigma_{\pi N}$. The enhancement
compared to the one loop chiral (or small scale) calculation can be
understood as follows. First, in the scalar ff of the pion one observes
sizeable two (and higher) loop corrections even at moderate $t$. This
is related to the strong pionic final--state interactions (FSI) in the isospin
zero S--wave. Similarly, the $\pi N$ partial wave $f_0^+ (t)$ exhibits
strong $\pi N$ FSI which can not be accounted for by a one--loop calculation.
Consequently, the corresponding spectral function shows a strong enhancement at the
two--pion threshold. This is at the heart of the large shift of the scalar
form factor. It is also worth pointing out that this does not mean that CHPT
breaks down. In fact, we are dealing with a small quantity which has a slowly
converging chiral expansion. Therefore, dispersion theory properly constrained
by chiral symmetry is the better tool to analyze this quantity (for details,
see ref.\cite{gls}).
Let me now turn to another process. The Mainz
group has investigated pion photoproduction in a dispersion theoretical
 framework.\cite{HDT}
While they did not enforce chiral constraints and did not include the data
very close to threshold in their analysis, the results for charged and neutral
pion production off the proton are very (even surprisingly) similar to 
the CHPT predictions.
In case of  $\pi^0$ production off the neutron, they find a sizeably
smaller electric dipole amplitude than the CHPT prediction. This might,
however, be an artefact of the not so precise data for this channel. It certainly
would be very interesting to combine their machinery with chiral symmetry
constraints. Clearly, more work in this direction should be performed. I
also would like to point out that for mesonic processes, the combination
of DR with CHPT has become a standard tool and it has been applied
successfully to many reactions, like e.g. $\pi \pi \to \pi\pi$, $\gamma\gamma
\to \pi^0 \pi^0$, $K^0 \to \pi^0 \gamma\gamma$, $\eta \to 3\pi$ and so on.

\subsection{Higher precision: Virtual photons and isospin violation}
Let me start with some general remarks. Since a large body of elastic
scattering and charge exchange data exists, one has the possibility of
deducing bounds on isospin violation from simple triangle relations,
which link e.g. the processes $\pi^\pm p \to \pi^\pm p$ and $\pi^- p
\to \pi^0 n$. Care has, however, to be taken since there are two
sources leading to isospin violation. One is the ``trivial'' fact that
electromagnetism does not conserve I--spin, since it couples to the charge.
The other one is a strong effect, linked to the difference of the light
quark masses, $m_d -m_u$. This is essentially the quantity one is
after. In terms of the symmetry breaking part of the QCD Hamiltonian,
i.e. the quark mass term, we have $
{\cal H}_{\rm QCD}^{\rm sb} = m_u \bar{u}u + m_d \bar{d}d
= \frac{1}{2}(m_u + m_d) ( \bar{u}u +  \bar{d}d ) +
\frac{1}{2}(m_u - m_d) ( \bar{u}u -  \bar{d}d )$,
so that the strong I--spin violation is entirely due to the isovector term.
This observation lead Weinberg\cite{mass} to address the question of I-spin
violation in the pion and the pion--nucleon sector, with the remarkable
conclusion that in neutral pion scattering off nucleons one should expect
gross violations of this symmetry, as large as 30\%. Only recently
an experimental proposal to measure the $\pi^0 p$ scattering length in
neutral pion photoproduction off protons below the secondary $\pi^+ n$
threshold has been presented and we are still far away from a determination
of this elusive quantity.\cite{aron}
At present, there exist two phenomenological analyses~\cite{gibbs,mats}
which indicate isospin breaking as large as 7\% in the S--waves (and smaller
in the P--waves). Both of these analyses employ approaches for the strong
interactions, which allow well to fit the existing data but can not  easily
be extended to the threshold or into the unphysical region. What is, however,
most disturbing is that the electromagnetic corrections have been calculated
using some prescriptions not necessarily consistent with the strong interaction
models used. One might therefore entertain the possibility that some of the
observed I--spin violation is caused by the mismatch between the treatment
of the em and strong contributions. Even if that is not the case, both models
do not offer any insight into the origin of the strong isospin violation, but
rather parametrize them. In CHPT, these principle problems can be circumvented
by constructing the most general effective Lagrangian with pions, nucleons
and virtual photons,\cite{ms}
\be
{\cal L}_{\pi N} =  {\cal L}_{\pi N}^{(1)} +  {\cal L}_{\pi N,{\rm str}}^{(2)} +
{\cal L}_{\pi N,{\rm em}}^{(2)} + {\cal L}_{\pi N,{\rm str}}^{(3)} +
{\cal L}_{\pi N,{\rm em}}^{(3)} + \ldots~.
\ee
where the superscript gives the chiral dimension. It is important to note that
the electric charge counts as a small momentum, based on the observation that
$e^2/4\pi \sim M_\pi^2 /(4\pi F_\pi )^2 \sim 1/100$. Since a virtual photon can
never leave a diagram, the local contact terms only start at dimension two.
For the construction of this effective Lagrangian and a more
detailed discussion of the various terms, see ref.\cite{ms}
Let me just review the results obtained so far (for more details,
see the talk by Steininger\cite{svent}). First, Weinberg's
finding concerning the scattering length difference $a(\pi^0 p) - a(\pi^0 n)$,
which is entirely given by a dimension two term $\sim m_u -m_d$, could be confirmed.
This is not surprising because for the case of neutral pions there is no contribution
from the em Lagrangian of order two and three. 
This will change at fourth order. Second, it was noted in\cite{ms}
that the I--spin breaking terms in $a(\pi^0 p)$ can be as large as the 
I--spin conserving ones.  It is also important
to note that the em effects are entirely due to the pion mass difference.
Another quantity of interest is the pion--nucleon $\sigma$--term. Of course,
one now has to differentiate between the one for the proton and the one for
the neutron, whose values to third order differ by the strong neutron--proton
mass difference. For the proton one finds
$\sigma_p(0) = \sigma_p^{IC} (0) + \sigma_p^{IV} (0) 
= 47.2~{\rm MeV} - 3.9~{\rm MeV}  = 43.3~{\rm MeV}$,
which means that the isospin--violating terms reduce the proton
$\sigma$--term by $\sim 8\%$. The electromagnetic effects are again
dominating the isospin violation since the strong contribution 
is just half of the strong
proton--neutron mass difference, 1~MeV. Furthermore, one gets
$\sigma_p (2M_{\pi^+}^2) - \sigma_p (0) = 7.5$~MeV, which 
differs from the result in the isospin limit at ${\cal O}(p^3)$ 
by a few percent. 
This small difference is well within the uncertainties related to the
so--called remainder at the Cheng--Dashen point.\cite{bkmcd}
Again, these corrections should be considered indicative since a fourth
order calculation is called for. As for the channels involving charged
pions, we have obtained some preliminary results. Considering only the
pion and nucleon mass differences (but neglecting all other virtual
photon effects), the CA prediction for the triangle--deviation (normalized
to the charge--exchange amplitude)
\beq
\Delta_{\pi N} = \frac{(T^{\rm EL+} - T^{\rm EL-}) - \sqrt{2}T^{\rm SCX}}
{ \sqrt{2}T^{\rm SCX}}~, 
\eeq
with $T^{\rm EL\pm}$ the elastic scattering amplitude $\pi^\pm p \to \pi^\pm p$
and $T^{\rm SCX}$ denotes the single charge exchange $\pi^-p \to \pi^0 n$, 
is 0.8\% (at threshold), and to second order we obtain 
$\Delta_{\pi N} = 2.3$\%. This number is smaller than the one obtained in
the phenomenological analyses\cite{mats,gibbs} but larger than expected by
many. Clearly, the one loop calculation has to be completed to draw
definite conclusions from this. Such an investigation is underway and the
details will be soon available.\cite{FMS2}
It is worth to emphasize again that we have a consistent machinery at hand to
simultaneously calculate the strong and the em isospin violating effects.

\section{Other topics}

Space does not allow to cover all the interesting developments in this
field since the last Baryons Conference in 1995. Still, I wish to briefly
discuss two topics, which have received lots of attention over the last
few years. Consider first the extension to {\it three} flavors.
While that is technically straightforward, the larger kaon mass and the
appearance of (subthreshold) resonances in certain channels clearly limit
the usefulness of baryon CHPT. However, for particular reactions and processes,
stringent predictions can be made. As an example, I mention the recent bounds on
the strange magnetic form factor of the nucleon, which will be of use in
analyzing the data from SAMPLE and CEBAF to constrain the strange magnetic
moment.\cite{HMS} Recently, a cut--off scheme was proposed to improve the
convergence properties of SU(3) baryon CHPT.\cite{DHB} It is argued that the
natural extension of the hadrons suppresses the unwanted short--distance
(non--chiral) physics. Technically, this is implemented by a cut--off which essentially
decreases the large kaon and eta loop contributions. To my opinion, 
in its present formulation this is
not a viable alternative since one shuffles the large loop contributions
into a string of large higher dimension terms generated through the cut--off
procedure. Certainly, this deserves more study.  Thirdly, there has also
been made considerable progress in setting up non--perturbative resummation
schemes to deal with the abovementioned resonances, see e.g.\cite{TUM,Val}
With a few parameters, one is  able to describe a wealth of data for a 
large energy range.  It should be noted, however, that such approaches seem to be
more useful for studying the
nature of some resonances (or quasi bound--states) than for directly testing 
the consequences of chiral symmetry breaking. 
Another subject of much discussion is the extension of EFT approaches
to few--nucleon systems. This is reviewed by David Kaplan.\cite{david}
I only wish to add that one still has to study in more detail alternative
schemes like e.g. the hybrid approach proposed by Weinberg,\cite{weinh} in
which nuclear matrix elements are calculated using realistic wave functions
and applying CHPT constraints to the interaction kernel. This approach has
been applied quite succesfully to neutral pion photoproduction off 
deuterium\cite{silas} and to $\pi-d$ scattering.\cite{silas2} 

Chiral nucleon dynamics has matured considerably over the last few years
and many more precise data are becoming available. Furthermore, there is
progress in extending the framework to deal with higher energies, the
three flavor case and few--nucleon systems.

\section*{References} 


\begin{thebibliography}{99}
\bibitem{wein79} S. Weinberg, {\it Physica} A {\bf 96}, 327 (1979).
\bibitem{bkmrev} V. Bernard, N. Kaiser and Ulf-G. Mei{\ss}ner,
{\it  Int.J. Mod. Phys.} E {\bf 4}, 193 (1995).
\bibitem{FMS}N. Fettes, Ulf-G. Mei{\ss}ner and S. Steininger,
\Journal{\NPA}{640}{199}{1998}.
\bibitem{bkmppn} V. Bernard, N. Kaiser and Ulf-G. Mei{\ss}ner,
\Journal{\NPA}{619}{261}{1997}.
\bibitem{MS} Martin Sevior,   {\it these proceedings}.
\bibitem{bkmpi0l} V. Bernard, N. Kaiser and Ulf-G. Mei{\ss}ner,
\Journal{\PLB}{378}{337}{1996}.
\bibitem{gerulf}G. Ecker and Ulf-G. Mei{\ss}ner, {\it Comm. Nucl. Part.
Phys.} {\bf 21}, 347 (1995).
\bibitem{TW} Th. Walcher, {\it these proceedings}.
\bibitem{silas}S.R. Beane, V. Bernard, T.-S.H. Lee, Ulf-G. Mei{\ss}ner
and U. van Kolck, \Journal{\NPA}{618}{381}{1997}.
\bibitem{berg} J.C. Bergstrom et al., \Journal{\PRC}{57}{3202}{1998}. 
\bibitem{SU}S. Steininger and Ulf-G. Mei{\ss}ner, 
\Journal{\PLB}{391}{446}{1997}.
\bibitem{TRImu}G. Jonkmans et al., \Journal{\PRL}{77}{4512}{1996}.
\bibitem{bkmgp} V. Bernard, N. Kaiser and Ulf-G. Mei{\ss}ner,
\Journal{\PRD}{50}{6899}{1994}.
\bibitem{THp}T.R. Hemmert, {\it these proceedings}.
\bibitem{Ecker}G. Ecker,  {\tt  hep-ph/9805500}.
\bibitem{ulfhugs} Ulf-G. Mei{\ss}ner,  {\tt  hep-ph/9711365}.
\bibitem{bkkm}V. Bernard, N. Kaiser, J. Kambor and Ulf-G. Mei{\ss}ner,
\Journal{\NPB}{388}{315}{1992}.
\bibitem{Th1}T.R. Hemmert, B.R. Holstein, J. Kambor and G. Kn\"ochlein,
\Journal{\PRD}{57}{5746}{1998}.
\bibitem{nicole} N. d'Hose, {\it these proceedings}.
\bibitem{Th2}T.R. Hemmert, B.R. Holstein, G. Kn\"ochlein and S. Scherer,
\Journal{\PRD}{55}{2630}{1997}.
\bibitem{Th3}T.R. Hemmert, B.R. Holstein, G. Kn\"ochlein and S. Scherer,
\Journal{\PRL}{79}{22}{1997}.
\bibitem{Th4} D. Drechsel, T.R. Hemmert et al., in preparation. 
\bibitem{bkmlet}V. Bernard, N. Kaiser and Ulf-G. Mei{\ss}ner,
\Journal{\PRL}{74}{3752}{1995}.
\bibitem{jmd} E. Jenkins and A. Manohar, \Journal{\PLB}{259}{353}{1991}.
\bibitem{HHK} T.R. Hemmert, B.R. Holstein and J. Kambor,
\Journal{\PLB}{395}{89}{1997}; 
{\tt  hep-ph/9712496}, {\it J. Phys.} G, in print.
\bibitem{bkmzm} V. Bernard, N. Kaiser and Ulf-G. Mei{\ss}ner, 
{\it Z.  Phys.} C  {\bf 60}, 111 (1993). 
\bibitem{JK}J. Kambor, {\tt  hep-ph/9711484}. 
\bibitem{gls} J. Gasser, H. Leutwyler and M.E. Sainio,
\Journal{\PLB}{253}{252}{1991}.
\bibitem{BFHM} V. Bernard et al., 
\Journal{\NPA}{635}{121}{1998}.
\bibitem{GC} G. Gellas, {\it these proceedings}.
\bibitem{BHM} V. Bernard, T.R. Hemmert and Ulf-G. Mei{\ss}ner,
{\it in preparation}.
\bibitem{tnt}T.N. Truong, \Journal{\PRL}{61}{2526}{1988}. 
\bibitem{gm}J. Gasser and Ulf-G. Mei{\ss}ner, \Journal{\NPB}{357}{90}{1991}.
\bibitem{JD}J.F. Donoghue, {\tt hep-ph/9607351}. 
\bibitem{HDT}O. Hanstein, D. Drechsel and L. Tiator, \Journal{\PLB}{399}{13}{1997};
\Journal{\NPA}{632}{591}{1998}.
\bibitem{mass}S. Weinberg, Trans. N.Y. Acad. of Sci. 38, 185 (1977).
\bibitem{aron}A.M. Bernstein, $\pi$N Newsletter {\bf 9}, 55 (1993).
\bibitem{gibbs}W.R. Gibbs, Li Ai and W.B. Kaufmann, \Journal{\PRL}{74}{3740}{1995}.
\bibitem{mats}E. Matsinos, \Journal{\PRC}{58}{3014}{1997}.
\bibitem{ms}Ulf-G. Mei{\ss}ner and S. Steininger,
\Journal{\PLB}{419}{403}{1998}.
\bibitem{svent} S. Steininger, {\it these proceedings}.
\bibitem{bkmcd} V. Bernard, N. Kaiser and Ulf-G. Mei{\ss}ner,
\Journal{\PLB}{389}{144}{1996}.
\bibitem{FMS2}N. Fettes, Ulf-G. Mei{\ss}ner and S. Steininger,  {\it in preparation}.
\bibitem{HMS}T.R. Hemmert, Ulf-G. Meissner and S. Steininger, 
{\tt hep-ph/9806226}, {\it Phys. Lett.} {\bf B}, in print. 
\bibitem{DHB}J. F. Donoghue, B. R. Holstein and B. Borasoy, {\tt hep-ph/9804281}. 
\bibitem{TUM}N. Kaiser, P.B.Siegel and W. Weise,
\Journal{\NPA}{594}{325}{1995};\\ N. Kaiser, T. Waas and W. Weise, 
\Journal{\NPA}{612}{297}{1997}.
\bibitem{Val}E. Oset and A. Ramos, \Journal{\NPA}{635}{99}{1998}.   
\bibitem{david} D.B. Kaplan, {\it these proceedings}.
\bibitem{weinh} S. Weinberg, \Journal{\PLB}{295}{114}{1992}.
\bibitem{silas2} S.R. Beane, V. Bernard, T.-S.H. Lee, Ulf-G. Mei{\ss}ner,
\Journal{\PRC}{57}{424}{1998}.
\end{thebibliography}
\end{document}